\documentclass[letterpaper, 10 pt, conference]{ieeeconf}  

\IEEEoverridecommandlockouts                              

\usepackage{graphicx}
\usepackage{url}  
\usepackage{amsmath}
\usepackage{amssymb}
\usepackage{multirow}
\usepackage[table,xcdraw]{xcolor}

\title{\LARGE \bf
The Impact of Frequency Bands on Acoustic Anomaly Detection of Machines using Deep Learning Based Model
}

\author{Tin~Nguyen$^{1*}$, 
        Lam~Pham$^{2*}$,
        Phat~Lam$^{3}$,        
        Dat~Ngo$^{4}$,
        Hieu~Tang$^{5}$,
        Alexander~Schindler$^{6}$  
\thanks{L. Pham and A. Schindler are with Austrian Institute of Technology, Austria.}%
\thanks{T. Nguyen and P. Lam are with HCM University of Technology, Vietnam}%
\thanks{D. Ngo is with University of Essex, UK}%
\thanks{H. Tang is with FPT University, Vietnam}%
\thanks{(*) Main and equal contribution into the paper.}
}

\begin{document}

\maketitle
\thispagestyle{empty}
\pagestyle{empty}

\begin{abstract}
In this paper, we propose a deep learning based model for Acoustic Anomaly Detection of Machines, the task for detecting abnormal machines by analysing the machine sound.
By conducting extensive experiments, we indicate that multiple techniques of pseudo audios, audio segment, data augmentation, Mahalanobis distance, and narrow frequency bands, which mainly focus on feature engineering, are effective to enhance the system performance.
Among the evaluating techniques, the narrow frequency bands presents a significant impact.
Indeed, our proposed model, which focuses on the narrow frequency bands, outperforms the DCASE baseline on the benchmark dataset of DCASE 2022 Task 2 Development set. 
The important role of the narrow frequency bands indicated in this paper inspires the research community on the task of Acoustic Anomaly Detection of Machines to further investigate and propose novel network architectures focusing on the frequency bands.  

\end{abstract}
\section{INTRODUCTION}
\label{intro}
In industrial settings, the reliability of machinery is crucial for ensuring uninterrupted operations, which plays a decisive role in achieving production efficiency and cost-effectiveness. 
Therefore, machine condition monitoring, which involves the continuous monitoring of machinery to ensure machine reliability, is considered as an essential component in industrial settings. 
%
In terms of preventing machine failures, anomalies in machine sound are considered as one of the potential signals indicating the machinery breakdown situation. 
%
%
Therefore, the task of Acoustic Anomaly Detection of Machines (AADoM) proves valuable in the early detection of potential faults through sound observation, serving as the primary and proactive stage of an effective machine condition monitoring system. 
The AADoM has received attention from the scientific research community in recent years and become an increasingly challenging task in the widespread development stage of AI-based factory automation. 
%
%
%
Since the machine typically operates without issues for the majority of its runtime, collecting a number of anomalous sounds seems to be challenging and unrealistic. 
Therefore, the self-supervised/unsupervised approaches proposed for AADoM task have become popular.
Another challenge of the AADoM task is the issue of domain shifting when there are variations in acoustic characteristics between training and test data caused by some factors such as operational speed, different environmental settings, noise, etc. 
This leads models to fail to detect anomalies within different domains. 
For this reason, the pursuit of developing AADoM systems using domain generalization techniques has become a common objective in real-life applications. 
To deal with these two main challenges, a wide range of unsupervised/self-supervised and deep learning based approaches combining with domain generalization techniques have been proposed that successfully explore the sound features and distinguish normal or abnormal behavior of the machine sound. 
For example,~\cite{vae, dadaed, wang} utilized Autoencoder to learn the normal audio data from various domains and extract the latent space representing normal audio.  
Then, the anomaly score is calculated based on construction error between the evaluating audio data with the latent space. 
Another approach involves leveraging self-supervised based models that learn different attributes of available normal data across various domains such as machine operation conditions, machine types, noise conditions and use them as feature extractors to get audio embeddings in the representation of normal data. 
Some deep neural network based architectures are proposed to extract audio embeddings in a self-supervised manner such as EfficientNet-B0~\cite{twostage}, ResNet~\cite{wilkinghoffoutlier}, Efficient Residual Net~\cite{erran},  MobileNetV2~\cite{nejjar2022_mobilenet}, Mobile Facenet~\cite{zeng2022robust, morita2022comparative}, etc. 
Then, anomaly detectors such as Gaussian mixture models (GMM)~\cite{gauss_outlier}, local outlier factor (LOF)~\cite{lof}, or k-nearest neighbors (k-NN)  are manipulated to calculate anomaly score from the extracted audio embeddings using different metrics such as Euclidean distance or Cosine similarity. 
%

Generally, the existing AADoM systems mostly focus on three following strategies to enhance the model performance: improving network architectures to learn audio features better from given audio data across domains and then extract a generalized representation for normal audio data; improving anomaly detector algorithms and the estimated distribution of normal data to better differentiate anomalous samples among normal ones; improving domain generalization techniques to handle domain shifting problem. 
However, most of the existing AADoM systems have not focused on exploring feature extraction where spectrograms are generated and represent the audio input.
In particular, these systems used the entire spectrogram which provides a vast amount of information, but not all of the information in the spectrogram is necessarily indicative of anomalies. (i.e. white noise at low frequency, irrelevant spectral lines at very high frequency). 
Second, different types of machines exhibit distinctive characteristic sounds related to their components and mechanisms. The relevant patterns of normality and anomalies may be concentrated in specific frequency bands for a particular machine. 
Therefore, analyzing all frequency bands within the entire spectrogram for all machine types may overlook localized anomaly patterns, making it more challenging to detect and interpret insightful features. 
Finally, training a model on the entire spectrogram can be more complex, requiring a larger amount of labeled data and storage capacity, which is unsuitable for real-time applications or systems with limited computational resources. 
In this paper, we therefore focus on exploring feature extraction and then indicate the main factors influencing the model performance.
Initially, we propose a baseline system for the AADoM task based on the self-supervised and the deep neural network based approach.
To analyze feature extraction, we conducted extensive experiments to assess the 
influence of various factors of pseudo audios, audio segments, data augmentation and frequency bands.
The experimental results on the benchmark dataset of DCASE 2022 Task 2 Development set prove that focusing on the narrow frequency bands significantly facilitates AADoM task. 


\section{The Proposed Baseline System}
\label{our_system}
\begin{figure}[t]
    	\vspace{0.2cm}
    \centering
    \includegraphics[width =1\linewidth]{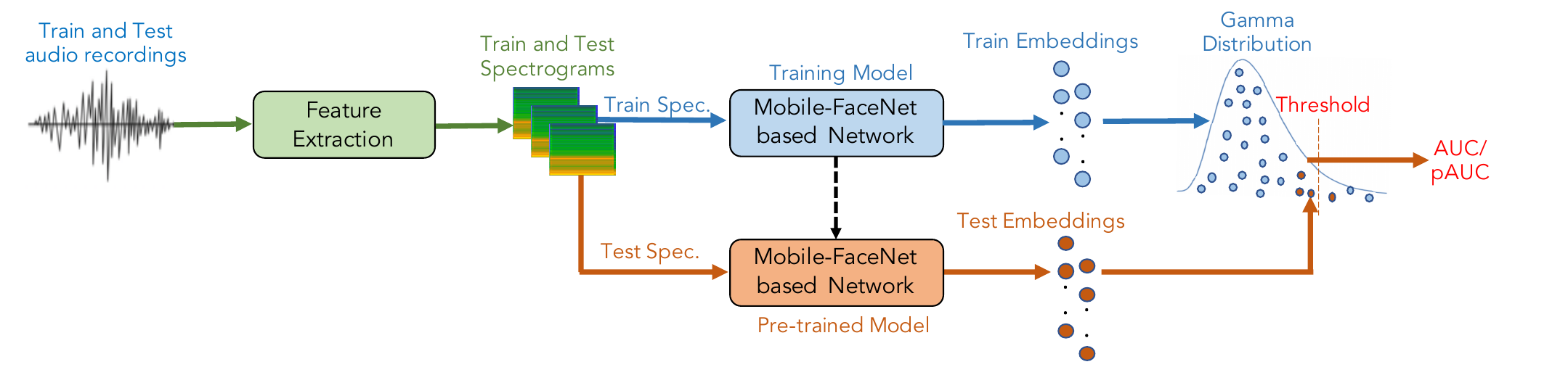}
   	\vspace{-0.5cm}
	\caption{The proposed baseline system for AADoM}
    	\vspace{-0.5cm}
    \label{fig:f1}
\end{figure}

We first propose a baseline system for the task of Acoustic Anomaly Detection of Machines (AADoM), referred to as the ProBaseline.
As Fig.~\ref{fig:f1} shows, the ProBaseline comprises three main components: feature extraction, classification model, Gamma distribution.

\textbf{Feature extraction:} The raw audio recordings from both `train' and `test' subsets are first transformed into STFT spectrograms with the window size and the hop size set to 2048 and 1024, respectively.
Then, we apply 128 Mel filters on STFT spectrograms to generate the Mel spectrograms.

\textbf{Classification Model:} Given the Mel-spectrograms represented the audio input, we establish the classification task in which Mel-spectrograms are classified into certain classes based on different operating conditions of the machines. 
In this paper, we proposed a Mobile-FaceNet based architecture for the classification task which is presented in Table~\ref{mface}.
Our proposed Mobile-FaceNet network is inspired by~\cite{mobilenetv2} which presents
an inverted residual structure with linear bottleneck. 
By leveraging this architecture, we add some layers such as LinearGDConv2d, LinearConv2d and Dense to create our deep learning model.
Notably, only audio data from the `train' subset is used for training the proposed Mobile-FaceNet network.

\textbf{Gamma Distribution:} After the training process, we achieve the pre-trained Mobile-FaceNet network.
We then feed the Mel-spectrograms from both the `train' and `test' subsets into the pre-trained network to extract the `train' audio embeddings and `test' audio embeddings, respectively. 
The `train' and `test' audio embeddings are the output of the Softmax layer in the pre-trained Mobile-FaceNet network. 
Given the `train' audio embeddings, we apply the Euclidean distance measurement~\cite{euclide}to compute the mean of the `train' audio embeddings.
We then establish the Gamma distribution in which the difference between one `train' audio embedding and the mean of the `train' audio embeddings is considered as one variable of the distribution, which is computed by:
\begin{equation}
    \label{eq:euclide_distance}
    d_e = ||\mathbf{x}-\mathbf{m}||^2_2
\end{equation}
where $||.||$ denotes 
Euclidean norm, $\mathbf{x}$ presents one `train' audio embedding, and $\mathbf{m}$ presents the mean of the `train' audio embeddings. 

For the evaluation process on the `test' subset, the difference between one `test' audio embedding and the mean of the `train' audio embeddings is first computed.
Then, the difference value is compared with the given Gamma distribution with a certain threshold (e.g. 0.9) to decide whether the `test' audio embedding is normal or abnormal.

\begin{table}[t]
\caption{The proposed Mobile-FaceNet based network architecture for the classification} 
       	\vspace{-0.2cm}
    \centering
    {
\begin{tabular}{|l|c|c|c|c|}
\hline
\textbf{Operations} & \textbf{t} & \textbf{c} & \textbf{n} & \textbf{s}                                                                  
\\  \hline
Conv2d [3x3]                  &  -       & 64     & -    & 2             \\ 
Conv2d [3x3]                  &  -       & 64     & -    & 1             \\ 
Bottleneck                  &  2       & 128    & 2    & 2             \\ 
Bottleneck                  &  4       & 128    & 2    & 2             \\ 
Bottleneck                  &  4       & 128    & 2    & 2             \\ 
Conv2d [1x1]                  &  -       & 512    & -    & 1             \\ 
LinearGDConv2d             &  -       & 512    & -    & 1             \\ 
LinearConv2d [1x1]            &  -       & 512    & -    & 1             \\ 
GlobalAvgPooling2d          &  -       & -      & -    & -             \\ 
Dense (activation='relu')    &  -       & 1024   & -    & -             \\ 
Dropout (0.3)                &  -       & -      & -    & -             \\
Dense (activation='softmax') &  -       & C      & -    & -             \\ \hline 
\end{tabular}
    }
   	\vspace{-0.5cm}
\label{mface}
\end{table}

\section{Further Improve the Proposed Baseline}
As the proposed baseline system (ProBaseline) is recently described, we assume that there are three main factors affecting the performance: the feature extraction to generate spectrogram input, the network architecture to extract audio embeddings, the distance measurement method to generate the Gamma distribution.  
In this paper, we focus on improving the quality of spectrogram input and the Gamma distribution rather than the network architecture.
We then propose our improvement methods which are described in below sections.

\subsection{Generate Pseudo Audio Samples}
We are inspired that if pseudo audio recordings can be generated and grouped into one class, the proposed Mobile-FaceNet based architecture is enforced to learn audio features and generate diverse audio embeddings. 
More diverse audio embeddings have the  potential to present a well-performed Gamma distribution that leads to improve the performance. 
In this paper, we therefore apply the pitch-shifting technique to synthesize pseudo audio recordings.

\subsection{Working on Audio Segment Level}
Instead of using the entire 10-second audio recording to generate one spectrogram and then extract one audio embedding, we split the entire audio recording into many audio segments with the length of 2.5 seconds per segment with an overlap of 50\%. 
By approaching the audio segments, we achieve a number of audio embeddings that enrich the Gamma distribution and then potentially further improve the AADoM performance.

\subsection{Apply Online Data Augmentations}
To enhance the Mobile-FaceNet network performance for the classification task and then generate well-performed audio embeddings, we apply two data augmentation methods, referred to as Specaugment~\cite{yusuf}, and Mixup~\cite{jia}. 
First, ten random and continuous temporal and frequency bins of the Mel-spectrograms are erased (Specaugment). 
Then, the spectrograms are randomly mixed together using different coefficients from both Beta and Uniform distributions (Mixup). 
As these two data augmentation methods are applied on each batch of Mel-spectrograms in the training process, we refer them to as the online data augmentations.
\begin{figure}[t]
    \centering
    \includegraphics[width =0.85\linewidth]{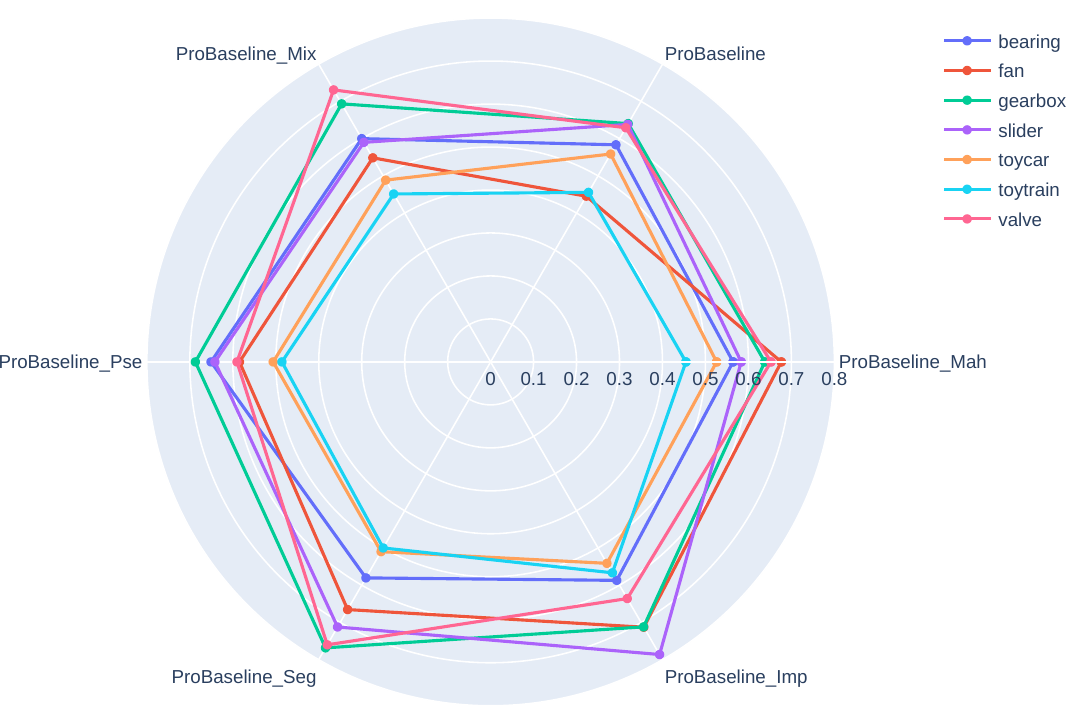}
       	\vspace{-0.3cm}
	\caption{Performance comparison (AUC) on \textbf{`source'} test domain among the proposed baseline (ProBaselline) and the proposed baseline with a certain improvement (ProBaseline-Mah, ProBaseline-Mix, ProBaseline-Pse, ProBaseline-Seg, and ProBaseline-Imp)}
    \label{fig:res_01}
\end{figure}
\subsection{Evaluate Narrow Frequency Bands}
We are inspired that certain faults in machine sound are located at certain frequency bands rather than spreading all the frequency bins.
Therefore, if we can pinpoint and evaluate certain frequency bands indicating the location of faults in machine sounds, the AADoM system is then designed to focus on these certain frequency bands for further enhancing the performance.
In this paper, we conduct an extensive experiment in which a wide range of frequency bands of 0 kHz to 3 kHz, 0.5 kHz to 3.5 kHz, 1 kHz to 4 kHz, 1.5 kHz to 4.5 kHz, 2 kHz to 5 kHz, 2.5 kHz to 5.5 kHz, 3 kHz to 6 kHz, 3.5 kHz to 6.5 kHz, 4 kHz to 7 kHz are investigated.
While evaluating these specific and narrow frequency bands, the other frequency bands on STFT spectrograms are removed before transforming into Mel-spectrograms.


\subsection{Apply Mahalanobis Distance}
As the Euclidean distance~\cite{euclide} used to measure the distance among audio embeddings is limited to present the correlation of the single variables in the audio embeddings (i.e. an audio embedding mathematically presents a vector with multiple variables)
Therefore, approaching the Euclidean can be distorted by outliers or skewed distributions as it treats variables of the audio embeddings independent and equal importance.
To tackle this limitation, we replace the Euclidean distance with the Mahalanobis distance measurement~\cite{maha}.
As the Mahalanobis method computes the mean of a multivariate distribution based on the covariance matrix of the distribution, it takes account of the scale, the correlation, and the shape of the variables.

The Mahalanobis distance between an audio embedding and a distribution of audio embeddings is calculated by the equation:

\begin{equation}
    \label{eq:maha_distance}
    d_m=\sqrt{(\mathbf{x}-\mathbf{m})^T\mathbf{S}^{-1}(\mathbf{x}-\mathbf{m})}
\end{equation}
where $\mathbf{x}$ is one audio embedding, $\mathbf{m}$ is the mean of the all audio embeddings, $\mathbf{S}$ is the covariance matrix of all audio embeddings.

\begin{figure}[t]
    \centering
    \includegraphics[width =0.85\linewidth]{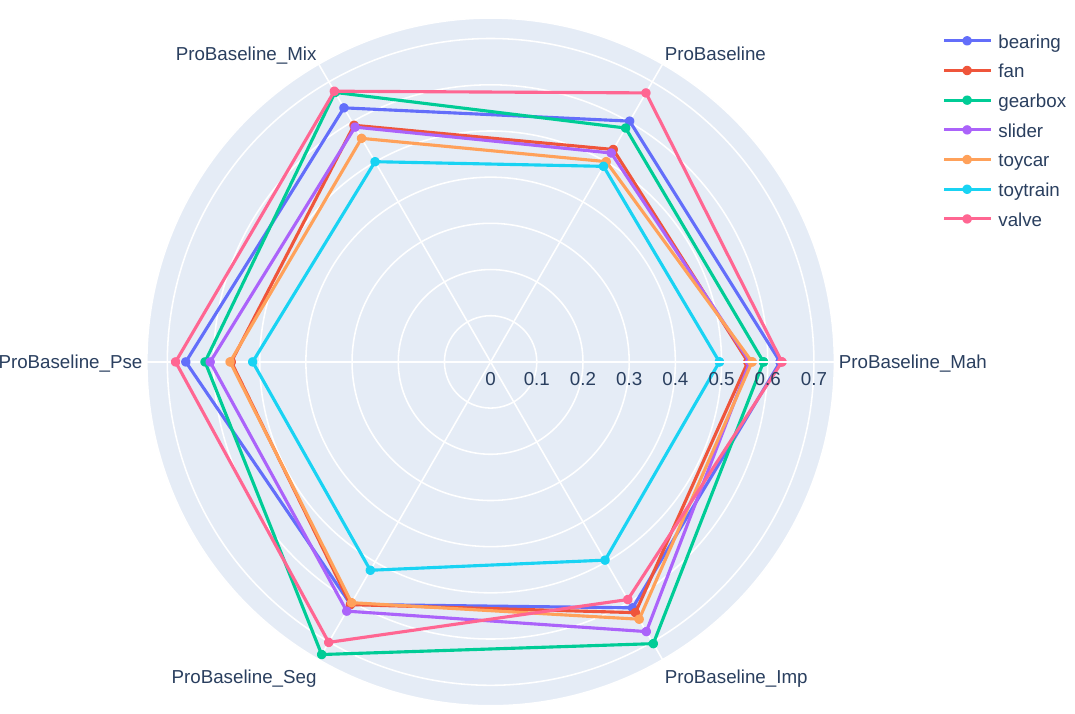}
       	\vspace{-0.3cm}
	\caption{Performance comparison (AUC) on \textbf{`target'} test domain among the proposed baseline (ProBaselline) and the proposed baseline with a certain improvement (ProBaseline-Mah, ProBaseline-Mix, ProBaseline-Pse, ProBaseline-Seg, and ProBaseline-Imp)}
    \label{fig:res_02}
\end{figure}


\subsection{Evaluate Different Thresholds of Gamma Distribution}
As the threshold value set to Gamma distribution may affect the performance of AADoM system, we evaluate a wide range of threshold values from 0.85 to 0.95 with the step of 0.1.

\section{Experiment And Results}
\subsection{Datasets}
In this paper, we evaluate our proposed models on the benchmark dataset: Development set of DCASE 2022 Challenge Task2. 
This dataset presents 10-second audio recordings which  were collected from 7 machine types: `Bearing', `Fan', `Gearbox', `Slider', `ToyCar', `ToyTrain' and `Valve'. 
For each machine, there are two data subsets of `train' and `test' for training and testing processes, respectively.
In the `train' subset, audio recordings are separated into three sections, referred to as `section 0', `section 1' and `section 2'. 
Three sections present three different operating conditions of the machine.
For each section, there are 1000 audio recordings which are separated into 990 `source' audio recordings and 10 `target' audio recordings. 
The imbalanced number between `source' and `target' data presents the issue of domain shift in this challenge.
As regards the `test' subset, it comprises 200 audio recordings which are separated into 100 `source' audio recordings and 100 `target' audio recordings, referred to as the `source' test domain and `target' test domain.  
\subsection{Evaluation Metric}
In this paper, we obey the DCASE 2022 Task 2 challenge~\cite{baseline_dcase,mimidg,toyadmos}, use AUC and pAUC as the evaluation metrics for the task of Anomalous Sound Detection on Machines.
%
\begin{figure}[t]
    \centering
    \includegraphics[width =0.9\linewidth]{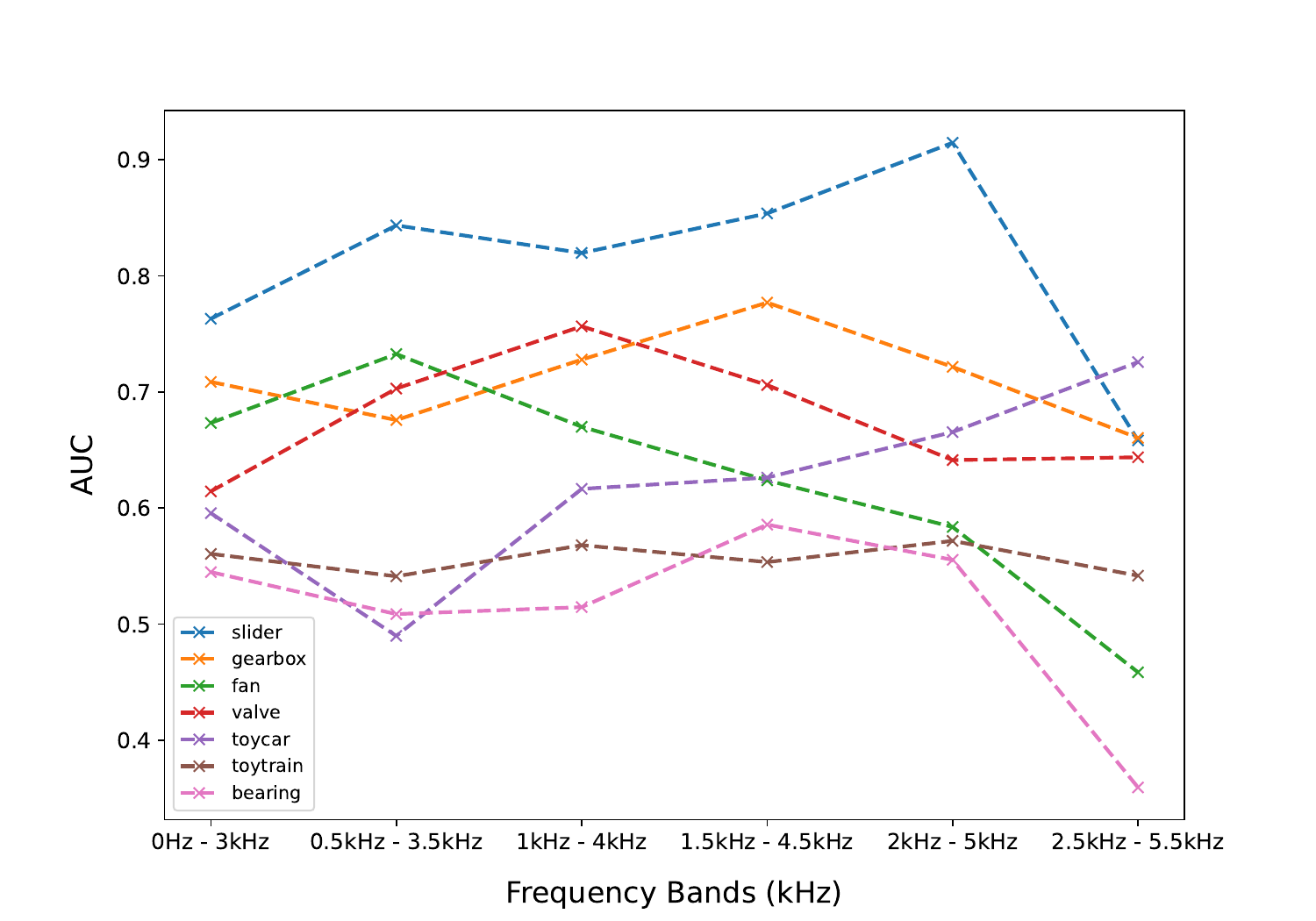}
           	\vspace{-0.3cm}
	\caption{The effect of frequency bands on \textbf{`source'} test domain}
       	\vspace{-0.4cm}
    \label{fig:res_03}
\end{figure}
\begin{figure}[t]
    \centering
    \includegraphics[width =0.9\linewidth]{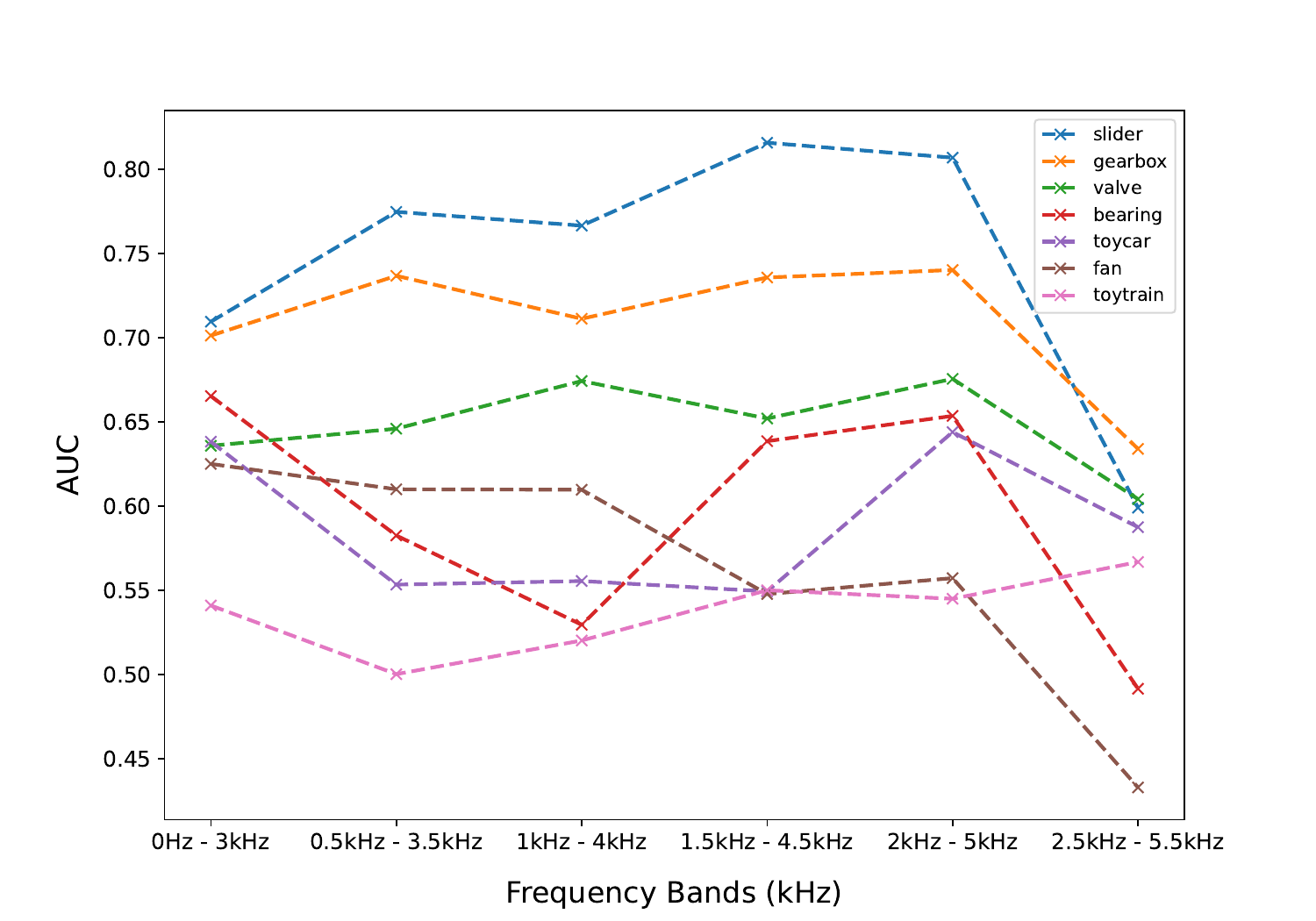}
       	\vspace{-0.3cm}
	\caption{The effect of frequency bands on \textbf{`target'} test domain}
            	\vspace{-0.3cm}
    \label{fig:res_04}
\end{figure}
\subsection{Experimental Settings}
We construct our proposed deep neural networks with the TensorFlow framework.
We train the proposed deep neural networks for 30 epochs.
All deep neural networks in this paper are trained with the Titan RTX 24GB GPU. 
We use the Adam method~\cite{Adam} for the optimization. 
The learning rate is set to 0.0001.
\subsection{Experimental Results and Discussion}
We first compare the proposed baseline (ProBaseline) to the proposed baseline with a certain improvement: the ProBaseline with Mahalanobis distance (ProBaseline-Mah), the ProBaseline with Mixup data augmentation (ProBaseline-Mix), the ProBaseline using audio segment (ProBaseline-Seg), the ProBaseline using pseudo samples (ProBaseline-Pse), and the ProBaseline using all improvements (ProBaseline-Imp).
\begin{figure}[t]
    \centering
    \includegraphics[width =0.9\linewidth]{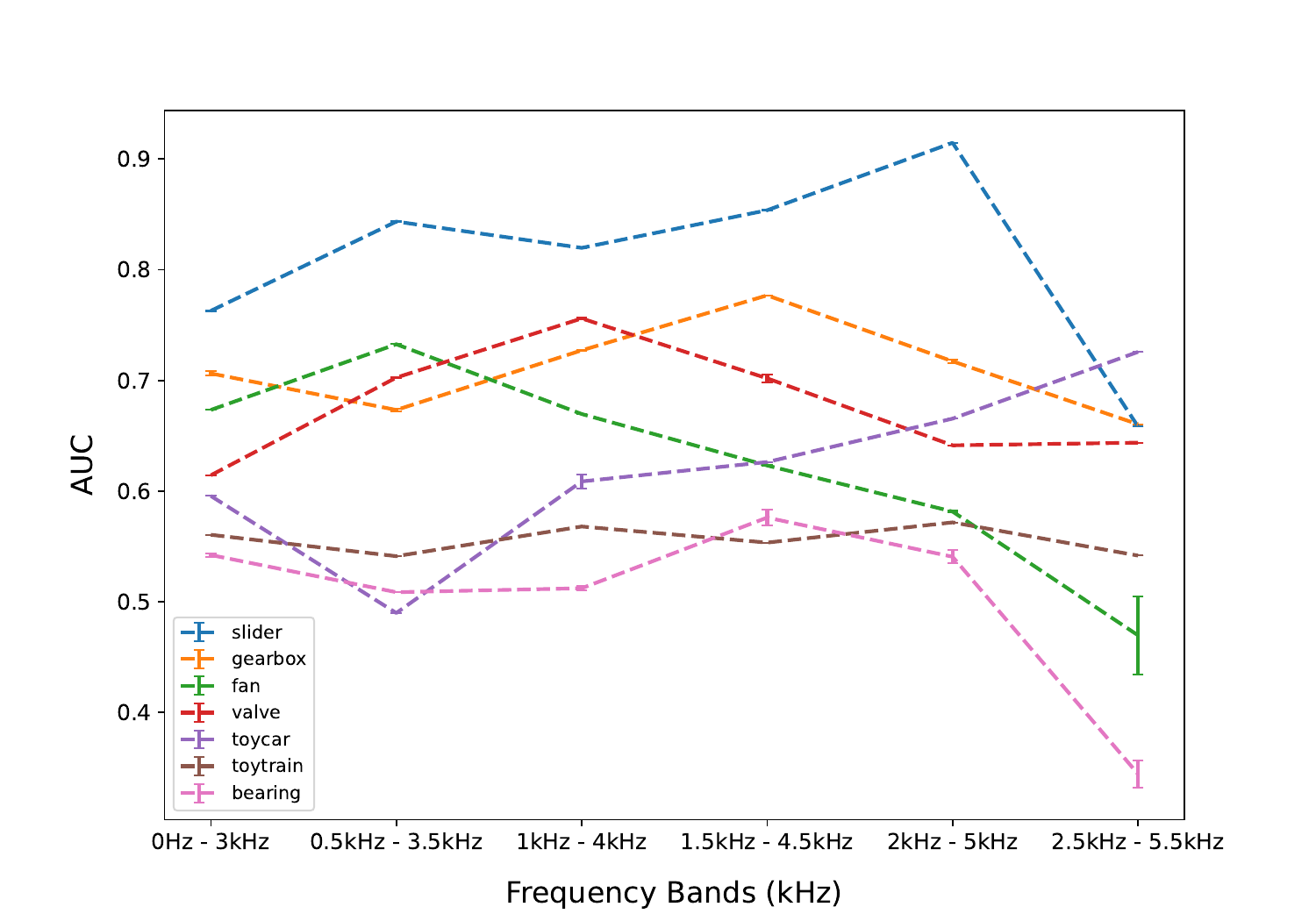}
               	\vspace{-0.3cm}
	\caption{The effect of Gamma threshold on \textbf{`source'} test domain}
        	\vspace{-0.4cm}
    \label{fig:res_05}
\end{figure}
\begin{figure}[t]
    \centering
    \includegraphics[width =0.9\linewidth]{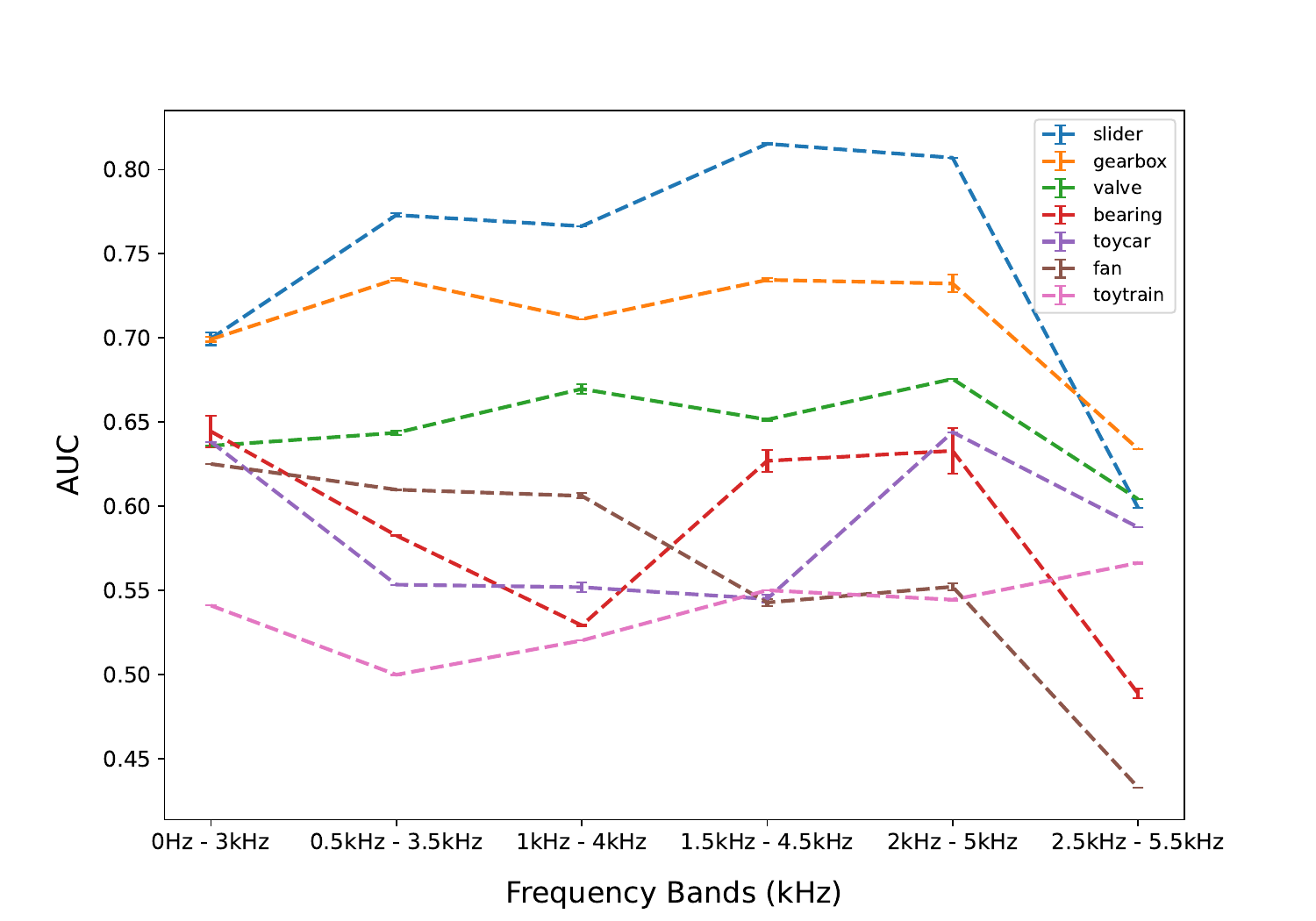}
       	\vspace{-0.3cm}
	\caption{The effect of Gamma threshold on \textbf{`target'} test domain}
            	\vspace{-0.3cm}
    \label{fig:res_06}
\end{figure}
As Fig.~\ref{fig:res_01} and Fig.~\ref{fig:res_02} show, all proposed improvement methods help to enhance the proposed baseline (ProBaseline) on all machines and for both `source' and `target' test generally.
Among improvement methods, using segments of spectrogram instead of an entire spectrogram significantly helps to improve the AUC scores (63,5\%/62.74\% from ProBaseline-Seg compared to 56.04\%/55.74\% from ProBasline in `source'/`test' domain).
This proves our motivation that using audio segments helps to generate more audio embeddings and create a well-performed Gamma distribution, then improve the AADoM system performance.
Given the proposed baseline using all improvement methods (ProBaseline-Imp), we evaluate whether narrow frequency bands affect the system performance.
We referred these models to as ProBaseline-Imp-Freq.
Notably, the Gamma distribution threshold is set to 0.9 to evaluate the specific frequency bands. 
As Fig.~\ref{fig:res_03} and Fig.~\ref{fig:res_04} show, the frequency bands significantly affect the performance regarding the type of machine.
In particular, while `Fan' and `Valve' machines present the best AUC of 73.2\%/61\% and 75.7\%/67.43\% on the `source'/`target' domain from the low bands of 0.5 kHz to 3.5 kHz and 1 kHz to 4 kHz, the `Slider' and `ToyCar' achieve the best AUC at the higher bands of 2 kHz to 5 kHz and 2.5 kHz to 5.5 kHz, respectively.
Both `Bearing' and `Gearbox' show the best AUC scores at the same middle bands of 1.5 kHz to 4.5 kHz.
Regarding the 'ToyTrain' machine, the frequency bands do not significantly affect the performance.
Notably, from 3 kHz to above, the performance significantly drops regarding all machine types. 
The experimental results indicate that faults occurring on different machine types are located at certain and narrow frequency bands. 
Therefore, if we focus on exploring frequency bands on which the corresponding faults of each machine locate, the AADoM system performance has the potential to be improved significantly.

Given the significant effect of frequency bands on the AADoM performance, we evaluate the role of Gamma distribution threshold over each frequency band.
In particular, we evaluate different threshold values from 0.85 to 0.95 with the step of 0.1.
As the Fig.~\ref{fig:res_05} and Fig.~\ref{fig:res_06} show, only `Bearing' and `Gearbox' machines are affected by the threshold at the frequency bands of 2 kHz - 5 kHz.
It can be concluded that the threshold does not significantly affect the AADoM system performance.

\begin{table}[t]
\caption{Performance comparison (AUC/pAUC) on `test' set among DCASE baseline, the proposed baseline (ProBaseline), the proposed baseline with all improvement methods (ProBaseline-Imp), and the proposed baseline with all improvement methods with focusing on frequency bands (ProBaseline-Imp-Freq) } 
       	\vspace{-0.2cm}
    \centering
    \scalebox{0.9}
    {
\begin{tabular}{|l|c|c|c|c|c|}
\hline
\textbf{Machine} & \textbf{DCASE} & \textbf{ProBaseline} & \textbf{ProBaseline} & \textbf{ProBaseline} \\ 
&\textbf{Baseline~\cite{baseline_dcase}} & &\textbf{-Imp} &\textbf{-Imp-Freq} \\
\hline
Bearing         &  60.2/57.1       & 59.2/51.3    & 60.1/54.4    & 63.1/56.9             \\ 
Fan             &  59.4/56.8       & 48.8/53.3    & 67.0/62.7    & 74.3/61.7             \\ 
Gearbox         &  62.7/56.0       & 61.2/56.6    & 70.8/62.4    & 75.5/64.6             \\ 
Slider          &  51.6/54.6       & 58.0/54.9    & 73.0/63.1    & 86.1/74.5             \\ 
ToyCar          &  55.5/52.2       & 52.9/51.1    & 59.2/54.2    & 65.6/56.1             \\ 
ToyTrain        &  51.5/51.5       & 47.2/50.1    & 53.1/52.4    & 55.7/52.5             \\ 
Valve           &  62.1/62.4       & 65.0/57.0    & 61.5/60.2    & 71.5/62.9             \\ 
\hline 
\end{tabular}
    }
  	\vspace{-0.4cm}
\label{final_res}
\end{table}
As the improvement methods and the focus on certain frequency bands help to improve the performance, we fine-tune the ProBaseline-Imp-Freq models by training the Mobile-FaceNet networks for more than 60 epochs. 
For each 20 epochs, we save the model and then extract audio embeddings.
The fine-tuning process helps to create a large number of audio embeddings that leads to achieve a well-performed Gamma distribution.
Our best results are then compared with the DCASE baseline in the Table~\ref{final_res}.
As the Table~\ref{final_res} shows, our fine-tuning ProBaseline-Imp-Freq model outperforms the DCASE baseline over all machines.
Among the machines, `ToyTrain' shows the worst performance AUC of 55.7\%.
Meanwhile, `Fan', `Gearbox', `Slide', and `Valve' present the potential AUC scores of 74.3\%, 75.5\%, 86.1\%, 71.5\%, respectively.

\section{Conclusion}
This paper has presented a deep learning based system for the Acoustic Anomaly Detection of Machines (AADoM).
By combining multiple techniques with a focus on feature extraction (pseudo audios, audio segment, data augmentation, frequency bands focusing), we successfully achieved an AADoM system that outperforms the DCASE baseline on the benchmark dataset of DCASE 2022 Task 2 Development set.
Our experimental results also indicate the significantly important role of narrow frequency bands in further improving the AADoM system performance.

\addtolength{\textheight}{-11cm}   

\bibliographystyle{IEEEbib}
\bibliography{refs}
\end{document}